\documentstyle[12pt]{article}

\newcommand{\text}{\rm}
\input{tcilatex}

\begin{document}

\title{{\bf Remarks on Infrared Dynamics in QED}$_{3}$}
\author{{\bf J. L. Boldo$^{\ast }$, B. M. Pimentel$^{\dagger }$ and J. L. Tomazelli$%
^{\dagger \dagger }$,}\vspace{2mm} \\
{\bf $^{\ast}$}CBPF, Centro Brasileiro{\bf \ }de Pesquisas F\'{\i}sicas \\
Rua Xavier Sigaud 150, 22290-180 Urca \\
Rio de Janeiro, Brazil\vspace{2mm}\\
\\
{\bf $^{\dagger }$}IFT, Instituto de F\'{\i}sica Te\'{o}rica\\
Rua Pamplona 145, 01405-900 Bela Vista\\
S\~{a}o Paulo, Brazil\vspace{2mm}\\
\\
{\bf $^{\dagger \dagger }$}DFQ - Faculdade de Engenharia, Universidade
Estadual Paulista\\
Campus de Guaratinguet\'{a}, Av. Dr. Ariberto Pereira da Cunha, 333\\
12500-000 Guaratinguet\'{a}\\
S\~{a}o Paulo, Brazil\vspace{2mm}}
\maketitle

\begin{abstract}
In this work we study how the infrared sector of the interaction Hamiltonian
can affect the construction of the S matrix operator of QED in (2+1)
dimensions.

\setcounter{page}{0}\thispagestyle{empty}
\end{abstract}


\vfill\newpage\ \makeatother
\renewcommand{\theequation}{\thesection.\arabic{equation}}

\section{\ Introduction\-}

The issue of infrared divergences in quantum electrodynamics has been better
understood since the discussion by Block and Nordsieck\cite{B-N}. They
pointed out the fact that the number of soft photons has no clear meaning
and is not an observable. In fact, in any scattering process involving
charged particles an infinite number of soft photons is emited collectively.
Lately, Yennie, Frautsch and Suura\cite{Yennie} showed that finite cross
sections from divergent S matrix elements are obtained if all radiative
corrections are taken into account.

An alternative approach to the problem using coherent states to represent
the asymptotic states with finite S matrix elements was proposed by Chung 
\cite{Chung}. He showed that all the S matrix elements can be made finite if
a suitable representation of \ photon states composed by coherent states is
chosen, rather than the usual Fock representation. Kulish and Faddeev \cite
{K-F} proposed to modify not only the space of asymptotic states but also to
redefine the scattering operator where the exact contribution of soft
photons is taken into account\cite{Kible, Zwanzinger}.

Interest in developing quantum electrodynamics in three dimensional
space-time is\ still growing. Although some lower dimensional models have
less physical content than those in four dimensions, they may provide new
insights, by avoiding many problems encountered in four dimensions, serving
as a theoretical laboratory. For instance, it is known that Einstein gravity
in four dimensional spacetime suffers from serious problems such as its
non-renormalizability ; however, the topological massive planar gravity is
unitary and renormalizable. In the same way, the treatment of infrared
divergencies in three dimensional gauge vector models can be performed by
adding a topological gauge invariant mass term in the Lagrangean, which
provides an infrared cutoff to the theory\cite{Jackiw, Deser}. Furthermore,
planar gauge theories have physical applications such as high-temperature
limits of four dimensional field theories and the quantum Hall effect.

Here we are concerned with the investigation of the infrared behaviour of QED%
$_{3}$, instead of calculating cross-sections explicitly. So we follow the
pioneering work of Murota\cite{Murota}, which sheds light into the
asymptotic limit of the theory. However, in contrast to the most of lower
dimensional problems which provide a simple setting where basic physical
phenomena can be worked out, these divergencies are more drastic in three
dimensions than in four and the interpretation of the results in the first
case is not so obvious. The basic idea is to consider the exact contribution
of the soft photons in the construction of the asymptotic time evolution
operator in the S operator definition.

In the next section we shall see that the soft photon contribution to the S
matrix may be isolated in a single time evolution operator, by making an
arbitrary separation of the interaction Hamiltonian. Section $3$ is devoted
to an examination of the physical content of this operator. The conclusions
are summarized in Section $4$.

\section{Preliminaries}

The starting point of the approach is to perform an arbitrary splitting of
the interaction Hamiltonian in H$_{s}$ , which contains soft photons, and H$%
_{h}$ the remaining one, i.e.,

\[
H(t)=H_s(t)+H_h(t). 
\]

Thus, the S matrix is given by

\begin{eqnarray}
S &=&T\exp \left[ -i\int_{-\infty }^{+\infty }dt\left( H_s\left( t\right)
+H_h\left( t\right) \right) \right]  \label{2.11} \\
&=&U\left( \infty ,t_0\right) S_hU\left( t_0,-\infty \right) ,  \nonumber
\end{eqnarray}
where 
\begin{equation}
U\left( t,t_0\right) =T\exp \left[ -i\int_{t_0}^td\tau H_s\left( \tau
\right) \right]  \label{2.12}
\end{equation}
and 
\begin{equation}
S_h=T\exp \left[ -i\int_{-\infty }^{+\infty }dtH_h\left( t\right) \right] .
\label{2.13}
\end{equation}

Note that the soft part of the Hamiltonian is all contained in the operator $%
U\left( t,t_{0}\right) $, so that its analysis will be the main purpose of
this paper. In addition, equation (\ref{2.11}) is of fundamental importance
because one stablishes the relation between the infrared (IR) behaviour and
asymptotic dynamics. In order to prove this relation we use the fact that $%
H_{s}(t)$ and $H_{h}(t)$ commute. Therefore,

\begin{eqnarray}
S &=&T\left\{ \exp \left[ -i\int_{-\infty }^{+\infty }dtH_{s}\left( t\right) %
\right] \left[ 1-i\int dt_{1}H_{h}\left( t_{1}\right) \right. \right. +
\label{2.18} \\
&&+\left. \left. \frac{\left( -i\right) ^{2}}{2!}\int dt_{1}\int
dt_{2}H_{h}\left( t_{1}\right) H_{h}\left( t_{2}\right) +...\right] \right\}
\nonumber \\
&=&T\left\{ U\left( +\infty ,-\infty \right) -i\int dt_{1}U\left( +\infty
,t_{1}\right) H_{h}\left( t_{1}\right) U\left( t_{1},-\infty \right) \right.
\nonumber \\
&&\left. +\frac{\left( -i\right) ^{2}}{2!}\int .dt_{1}\int dt_{2}U\left(
+\infty ,t_{1}\right) H_{h}\left( t_{1}\right) U\left( t_{1},t_{2}\right)
H_{h}\left( t_{2}\right) U\left( t_{2},-\infty \right) +...\right\} . 
\nonumber
\end{eqnarray}
In this way, the n$^{th}$ order term of $S$ matrix in the perturbation
series with respect to $H_{h}(t)$ is given by

\begin{eqnarray}
S^{\left( n\right) } &=&\left( -i\right) ^{n}\int dt_{1}...dt_{n}\theta
\left( t_{1}-t_{2}\right) \theta \left( t_{2}-t_{3}\right) ...\theta \left(
t_{n-1}-t_{n}\right)  \label{2.19} \\
&&\times U\left( +\infty ,t_{1}\right) H_{h}\left( t_{1}\right) U\left(
t_{1},t_{2}\right) H_{h}\left( t_{2}\right) ...U\left( t_{n-1},t_{n}\right)
H_{h}\left( t_{n}\right) U\left( t_{n},+\infty \right) .  \nonumber
\end{eqnarray}
Inserting complete sets before and after $H_{h}(t_{i})$, we see that the
dependence of $H_{h}(t_{i})$ is given by the difference between energies of
states before and after $t_{i}$

\[
H_{h}\left( t_{i}\right) \sim e^{i\Delta Et}. 
\]
On the other hand, we shall see that\ the$\ \ t_{i}$ dependence in the
operators $U\left( t_{n-i},t_{i}\right) $ and $U\left( t_{i},t_{n+i}\right) $
is as follows

\begin{equation}
U\left( t_{n-i},t_{i}\right) ,U\left( t_{i},t_{n+i}\right) \sim \sum
e^{i\sum \frac{p\cdot k}{E}t_{i}}.  \label{2.21}
\end{equation}
But, taking into account that the energy of soft photons is very small that
the hard ones, namely,

\begin{equation}
\left| \Delta E_i\right| \gg \left| \sum \frac{p\cdot k}{E_i}\right| ,
\label{2.22}
\end{equation}
we may use the approximation

\begin{equation}
U\left( t_{n-i},t_{i}\right) H_{h}\left( t_{i}\right) U\left(
t_{i},t_{n+i}\right) \simeq U\left( t_{n-i},t_{0}\right) H_{h}\left(
t_{i}\right) U\left( t_{0},t_{n+i}\right) .  \label{2.23}
\end{equation}
Since $U\left( t_{i},t_{0}\right) \simeq 1$, (\ref{2.19}) can be
approximated by

\[
S^{\left( n\right) }=U\left( \infty ,t_{0}\right) \frac{\left( -i\right) ^{n}%
}{n!}\int dt_{1}...dt_{n}T\left[ H_{h}\left( t_{1}\right) ...H_{h}\left(
t_{n}\right) \right] U\left( t_{0},-\infty \right) . 
\]
The above result holds for all n, and hence (\ref{2.11}) is proved.

\section{Construction of the Infrared Operator in QED$_3$}

We now shall construct $U\left( t,t_0\right) $ which contains the infrared
behaviour of the theory.

In the interaction picture, the QED Hamiltonian of the system of charged
particles and photons is

\begin{equation}
H(t)=\int d\stackrel{\rightarrow }{x}\Psi \left( x\right) \gamma _{\mu }%
\overline{\Psi }\left( x\right) \cdot A^{\mu }(x),  \label{2.25}
\end{equation}
where $\Psi $ and $\overline{\Psi }$ are the operators of the
electron-positron field and $A^{\mu }$ the operator of the electromagnetic
gauge field. These field operators can be expanded in terms of the creation
and annihilation operators as follows:

\[
A_{\mu }\left( x\right) =\frac{1}{2\pi }\int \frac{d\stackrel{\rightarrow }{k%
}}{\sqrt{2\omega _{k}}}\left( a_{\mu }\left( \stackrel{\rightarrow }{k}%
\right) e^{-ik\cdot x}+a_{\mu }^{\dagger }\left( \stackrel{\rightarrow }{k}%
\right) e^{ik\cdot x}\right) , 
\]

\begin{equation}
\Psi \left( x\right) =\Psi ^{\left( +\right) }\left( x\right) +\Psi ^{\left(
-\right) }\left( x\right) ,  \label{2.50}
\end{equation}

\begin{equation}
\overline{\Psi }\left( x\right) =\overline{\Psi }^{\left( +\right) }\left(
x\right) +\overline{\Psi }^{\left( -\right) }\left( x\right) ,  \label{2.51}
\end{equation}
where

\[
\Psi ^{\left( +\right) }\left( x\right) =\frac{1}{2\pi }\int d\stackrel{%
\rightarrow }{p}\sqrt{\frac{m}{p_{0}}}{\sum_{r}}b_{r}\left( \stackrel{%
\rightarrow }{p}\right) u_{r}\left( \stackrel{\rightarrow }{p}\right)
e^{-ip\cdot x}, 
\]

\begin{equation}
\Psi ^{\left( -\right) }\left( x\right) =\frac{1}{2\pi }\int d\stackrel{%
\rightarrow }{p}\sqrt{\frac{m}{p_{0}}}{\sum_{r}}d_{r}^{\dagger }\left( 
\stackrel{\rightarrow }{p}\right) v_{r}\left( \stackrel{\rightarrow }{p}%
\right) e^{ip\cdot x},  \label{2.52}
\end{equation}

\[
\overline{\Psi }^{\left( -\right) }\left( x\right) =\frac{1}{2\pi }\int d%
\stackrel{\rightarrow }{p}\sqrt{\frac{m}{p_{0}}}{\sum_{r}}b_{r}^{\dagger
}\left( \stackrel{\rightarrow }{p}\right) \overline{u}_{r}\left( \stackrel{%
\rightarrow }{p}\right) e^{ip\cdot x}, 
\]

\[
\overline{\Psi }^{\left( +\right) }\left( x\right) =\frac{1}{2\pi }\int d%
\stackrel{\rightarrow }{p}\sqrt{\frac{m}{p_{0}}}{\sum_{r}}d_{r}\left( 
\stackrel{\rightarrow }{p}\right) \overline{v}_{r}\left( \stackrel{%
\rightarrow }{p}\right) e^{-ip\cdot x}. 
\]
Here $a_{\mu }\left( \stackrel{\rightarrow }{k}\right) $, $b_{r}\left( 
\stackrel{\rightarrow }{p}\right) $ and $d_{r}\left( \stackrel{\rightarrow }{%
p}\right) $, are the annihilation operators of photon, electron and positron
respectively. $u_{r}\left( \stackrel{\rightarrow }{p}\right) $ and $%
v_{r}\left( \stackrel{\rightarrow }{p}\right) $ are the corresponding spinor
amplitudes. Inserting these quantities into (\ref{2.25}), one obtains for $%
H(t)$ an integral over the momenta $\stackrel{\rightarrow }{p}$, $\stackrel{%
\rightarrow }{q}$ and $\stackrel{\rightarrow }{k}$ of the fermions and
photons, which are related by the constraint 
\[
\stackrel{\rightarrow }{p}+\stackrel{\rightarrow }{q}=\stackrel{\rightarrow 
}{k}. 
\]

We now turn to investigate the infrared behaviour of the interaction
Hamiltonian, by making use of the following asymptotic approximations

\begin{equation}
p^0-q^0\mp \omega \simeq \mp \frac{p\cdot k}{p^0}.  \label{2.62}
\end{equation}

\[
p^0+q^0\pm \omega \simeq 2p^0 
\]
and

\begin{eqnarray}
\stackrel{-}{u}_r\left( p\right) \gamma ^\mu u_s\left( q\right) &=&\frac
1{2m}\stackrel{-}{u}_r\left( p\right) \left[ p^\mu +q^\nu +i\sigma ^{\mu \nu
}\left( p_\nu -q_\nu \right) \right] u_s\left( q\right)  \label{2.63} \\
&\simeq &\frac{p^\mu }m\delta _{rs},  \nonumber
\end{eqnarray}

It is easy to see that the time dependence of $H_{s}(t)$ split into two
groups, one is proportional to $e^{\pm 2ip_{o}t}$ and the other to $e^{\pm i%
\frac{k\cdot p}{p^{0}}t}$. Thus, after time integration, the former is
smaller than the latter due the fact that

\[
\left| p_{0}\right| \gg \left| \frac{p\cdot k}{p_{0}}\right| , 
\]
so that all those terms of the first group can be neglected.

It is important to notice that the vertex approximation (\ref{2.63}) is spin
independent; therefore, it is usually called eikonal in a sense that charged
particles act as pointlike sources.

Replacing these approximations into (\ref{2.25}) one finally obtains the
interaction Hamiltonian, which describes the interaction between charged
particles and soft photons:

\begin{equation}
H_s\left( t\right) =\frac 1{\left( 2\pi \right) }\int d\stackrel{\rightarrow 
}{k}\int d\stackrel{\rightarrow }{p}\frac 1{\sqrt{2\omega }}\frac{p^\mu }{p_0%
}\rho \left( \stackrel{\rightarrow }{p}\right) \left[ a_\mu \left( \stackrel{%
\rightarrow }{k}\right) e^{-i\frac{k\cdot p}{p^0}t}+a_\mu ^{\dagger }\left( 
\stackrel{\rightarrow }{k}\right) e^{i\frac{k\cdot p}{p^0}t}\right] ,
\label{2.64}
\end{equation}
where $\rho \left( \stackrel{\rightarrow }{p}\right) $ is the charge density
operator

\begin{equation}
\rho \left( \stackrel{\rightarrow }{p}\right) =e\left[ b_r^{\dagger }\left( 
\stackrel{\rightarrow }{p}\right) b_r\left( \stackrel{\rightarrow }{p}%
\right) -d_r^{\dagger }\left( \stackrel{\rightarrow }{p}\right) d_r\left( 
\stackrel{\rightarrow }{p}\right) \right] .  \label{2.66}
\end{equation}

Substituting (\ref{2.64}) in (\ref{2.12}) we obtain $U\left( t,t_{0}\right) $
in form of\ a T-ordered product. This expression can be transformated in to
an ordinary product by noting that $H_{s}\left( t\right) $ possesses the
following property

\begin{equation}
\left[ H_{s}\left( t_{1}\right) ,H_{s}\left( t_{2}\right) \right]
=c-number,\,\,\,\forall \,\,t_{1}\neq t_{2},  \label{2.14}
\end{equation}
which commutes with $H_{s}\left( t\right) $ for all $t,t_{1},t_{2}.$ By
making use of this commutation relation, one readily obtain a more suitable
form for the time evolution operator 
\begin{equation}
U\left( t,t_{0}\right) =\exp \left\{ R\left( t,t_{0}\right) \right\} \exp
\left\{ i\phi \left( t,t_{0}\right) \right\} ,  \label{2.15}
\end{equation}
where 
\begin{equation}
R\left( t,t_{0}\right) =\exp \left[ -i\int_{t_{0}}^{t}d\tau H_{s}\left( \tau
\right) \right]  \label{2.16}
\end{equation}
and 
\begin{equation}
i\phi \left( t,t_{0}\right) =\exp \left\{ -\frac{1}{2}\int_{t_{0}}^{t}d\tau
\int_{t_{0}}^{\tau }ds\left[ H_{s}\left( \tau \right) ,H_{s}\left( s\right) %
\right] \right\} .  \label{2.17}
\end{equation}

The $S$ matrix is defined in (\ref{2.11}) in the limit $t,s\rightarrow \pm
\infty .$ In order to simplify the calculations we shall take $t_0=0$, since
the time $t_0$ is a finite parameter chosen between $-\infty <t_0<+\infty $ $%
.$

Then, the construction of the asymptotic part of the $S$ matrix is made by
replacing these limits in the operators $U\left( t,t_{0}\right) $, namely,

\begin{equation}
U\left( \pm \infty ,0\right) =\exp \left[ R\left( \pm \infty ,0\right) %
\right] \exp \left[ i\phi \left( \pm \infty ,0\right) \right] ,
\label{2.113}
\end{equation}
where 
\begin{equation}
R\left( \pm \infty ,0\right) =\frac{1}{\left( 2\pi \right) }\int \frac{d%
\stackrel{\rightarrow }{k}}{\sqrt{2\omega }}\int d\stackrel{\rightarrow }{p}%
\frac{p^{\mu }}{p\cdot k}\rho \left( \stackrel{\rightarrow }{p}\right) \left[
a_{\mu }\left( \stackrel{\rightarrow }{k}\right) -a_{\mu }^{\dagger }\left( 
\stackrel{\rightarrow }{k}\right) \right]   \label{2.114}
\end{equation}
and 
\begin{equation}
\phi \left( \pm \infty ,0\right) =\mp \frac{i}{4\pi }\int d\stackrel{%
\rightarrow }{p}\int d\stackrel{\rightarrow }{q}\widehat{p}\cdot \widehat{q}%
\rho \left( \stackrel{\rightarrow }{p}\right) \rho \left( \stackrel{%
\rightarrow }{q}\right) \int \frac{d\stackrel{\rightarrow }{k}}{2\omega }%
\frac{\delta \left[ k\cdot \left( \widehat{p}-\widehat{q}\right) \right] }{%
\left( \widehat{q}\cdot k\right) }.  \label{2.116}
\end{equation}
Here

\[
\widehat{p}\equiv \frac p{p_0},\,\,\widehat{q}\equiv \frac q{q_0}. 
\]

The infrared behaviour of the phase operator can be most easily determined
in the nonrelativistic limit, $\left| \stackrel{\rightarrow }{v}\right| \ll
c $. In this case, it can be rewritten as

\begin{equation}
\phi \left( \pm \infty ,0\right) =\pm \frac{i}{4\pi }\int d\stackrel{%
\rightarrow }{p}\int dq\rho \left( \stackrel{\rightarrow }{p}\right)
v_{pq}\rho \left( \stackrel{\rightarrow }{q}\right) \left( -\frac{1}{\omega }%
\right) _{\lambda }^{\Delta },  \label{2.117}
\end{equation}
where the cutoff parameter $\Delta $ was used to define\ the infrared sector
and $\lambda \rightarrow 0.$ Also, 
\[
v_{pq}\equiv \frac{p\cdot q}{\sqrt{\left( p\cdot q\right) ^{2}-p^{2}q^{2}}},
\]
is usually called phase operator.

\section{Discussion}

From the above result, we see \ that the I.R. structure of the theory is
made explicit. Furthermore, it is instructive to compare the I.R. behaviour
of the phase operator (\ref{2.117}) in 3D and that one in 4D$.$ In the
latter case, the divergent Coulomb phase as a function of the photon
frequency does have the same logarithm behaviour as that one when the same
infrared analysis is made in terms of the asymptotic times\cite{Murota,
Ciafaloni}, while it is\ not true in 3D\cite{PTB}.

Moreover, the operators $R\left( \pm \infty ,0\right) $ are not well defined
in the Fock space for photons and charged particles. In fact, when applied
on states which belong to the Fock space, they yield coherent states. In
other words, they force us to introduce a new space of asymptotic states,
the space of coherent states, which contains a infinite number of soft
quanta and, therefore, becomes necessary to sum over final and initial
states and not only over final states as claimed by the Block-Nordsieck
theory.

\vspace{5mm}

{\Large {\bf Acknowledgements}}

J.L.B. thanks CNP$q$ \ for financial support; B.M.P. and J.L.T. are
partially supported by CNP$q$.


\vspace{5mm}

\begin{thebibliography}{99}
\bibitem{B-N}  F. Bloch and A. Nordsieck, Phys. Rev. {\bf 52}, 54, (1937);

\bibitem{Yennie}  D. R. Yennie, S.C. Frautschi, and H. Suura, Ann. Phys.
(New York), {\bf 13}, 379 (1961);

\bibitem{Chung}  V. Chung, Phys. Rev. {\bf 140B}, 1100 (1965);

\bibitem{K-F}  P. Kulish and L. Faddeev, Teor. Mat. Fiz. {\bf 4}, 153 (1970)
[Theor. Math. Phys. {\bf 4}, 745 (1970)];

\bibitem{Kible}  T. W. B. Kibble, Phys. Rev., {\bf 173}, 1527; {\bf 173},
1882; {\bf 175}, 1624 (1968);

\bibitem{Zwanzinger}  D. Zwanziger, Phys. Rev. {\bf D11}, 3481, 3504 (1975);

\bibitem{Jackiw}  S. Deser, R. Jackiw, and A. Templeton, Ann. Phys. {\bf 140}%
, 372 (1982); Phys. Rev. Lett. {\bf 48}, 975 (1982);

\bibitem{Deser}  S. Deser and Z. Yang, Class. Quantum Grav. {\bf 7}, 1603
(1990);

\bibitem{Murota}  T. Murota, Prog. Theor. Phys. {\bf 24}, 1109, (1960);

\bibitem{Ciafaloni}  M. Ciafaloni, Phys. Lett. {\bf 150B}, 379 (1985); S.
Catani, M. Ciafaloni and G. Marchesini, Nucl. Phys. B264, 588 (1986); S.
Catani, M. Ciafaloni, Nucl. Phys. {\bf B289}, 535 (1987);

\bibitem{PTB}  J. L. Boldo, B. M. Pimentel and J. L. Tomazelli, Can. J.
Phys. {\bf 76}, 69 (1998).
\end{thebibliography}
\end{document}